\documentclass[prl,reprint,amsmath,nofootinbib,superscriptaddress,showpacs]{revtex4-1}
\usepackage{amssymb}
\usepackage{empheq}

\usepackage[utf8]{inputenc}  
\usepackage{amsmath}    
\usepackage{amsfonts}
\usepackage{mathrsfs} 
\usepackage[unicode]{hyperref}
\usepackage{graphicx}
\newcommand{\beq}{\begin{equation}}
\newcommand{\eeq}{\end{equation}}
\newcommand{\bea}{\begin{eqnarray}}
\newcommand{\eea}{\end{eqnarray}}

\begin{document}

\begin{abstract}

An intriguing possibility that can address pathologies in both early universe cosmology (i.e. the horizon problem) and quantum gravity (i.e. non-renormalizability), is that particles at very high energies and/or temperatures could propagate arbitrarily fast. A concrete realization of this possibility for the early universe is the Tachyacoustic (or Speedy Sound) cosmology, which could also produce a scale-invariant spectrum for scalar cosmological perturbations. Here, we study Thermal Tachyacoustic Cosmology (TTC), i.e. this scenario with thermal initial conditions. We find that a phase transition in the early universe, around the scale of Grand Unified Theories (GUT scale; $T\sim 10^{15}$ GeV), during which the speed of sound drops by $25$ orders of magnitude within a Hubble time, can fit current CMB observations. We further discuss how production of primordial black holes constrains the cosmological acoustic history, while coupling TTC to Horava-Lifshitz gravity leads to a lower limit on the amplitude of tensor modes ($r \gtrsim 10^{-3}$), that are detectable by CMBpol (and might have already been seen by the BICEP-Keck collaboration). 

\end{abstract} 

\title{Thermal Tachyacoustic Cosmology}
\author{Abhineet Agarwal}                                                                                              
\affiliation{Department of Physics and Astronomy, University of Waterloo, Waterloo, ON, N2L 3G1, Canada}
\author{Niayesh Afshordi}
\email{nafshordi@pitp.ca}
\affiliation{Perimeter Institute for Theoretical Physics, 31 Caroline St. N., Waterloo, ON, N2L 2Y5, Canada}
\affiliation{Department of Physics and Astronomy, University of Waterloo, Waterloo, ON, N2L 3G1, Canada}

\date{\today}    

\maketitle

Our universe appears to be nearly homogenous on scales larger than tens of megaprsecs, and yet contains a variety of structures on essentially all observable scales. This structure is a result of small inhomogeneities in cosmic initial conditions, which are ubiquitous in cosmic microwave background (CMB) anisotropies, most recently, and comprehensively, mapped by the ESA's Planck satellite \cite{Ade:2013ktc}.  Planck (+WMAP)
observations \cite{Ade:2013zuv} indicate that these inhomogeneities are well-described by a
near-scale-invariant spectrum of adiabatic perturbations with the power spectrum: \bea
 {\cal P}_{\zeta} (k) = &&(2.196 \pm 0.059) \times
10^{-9} \nonumber\\ && \times\left ( k \over 0.05~{\rm Mpc}^{-1} \right)^{-0.0397 \pm
0.0073}, \label{p_zeta} \eea where $k$ is the comoving wavenumber
for spatial fluctuations.

So, how do different causally disconnected regions in the universe come into thermal equilibrium, with correlated, near-scale-invariant fluctuations?

 In \cite{AlanGuth}, it was proposed that the early universe has undergone a phase of rapid exponential expansion or ``inflation''. This could potentially solve many problems in big bang cosmology, including the formation of the large scale structure of the Universe and CMB anisotropies. Quantum fluctuations, which start in their ground state, cross the so-called Hubble horizon during inflation \cite{Mukhanov:1990me}, after which they stop their oscillatory behaviour and grow into large observable amplitudes. The Hubble horizon is crossed once $k c< aH$, where $c$ is the speed of light, $a$ is the cosmological scale factor, and $H=\dot{a}/a$ is the Hubble expansion rate. We can choose units where $c = 1$ and hence the above condition would reduce to 
$k < aH$.  At the end of inflation, the quantum fluctuations are imprinted into the plasma density perturbations, hence laying the seed for the formation of the large, and small, scale structure of the universe. 

In the standard inflationary paradigm introduced  in \cite{AlanGuth}, inflation was driven by a potential term of a scalar field. An alternative approach was developed in \cite{GarrigaandMukhanov,kinflation}, the so-called k-inflation,  which is driven by the non minimal terms in the kinetic energy. Unlike in the case of the standard potential driven inflation, the quantum fluctuations here freeze after they cross the {\it acoustic} horizon, rather than the Hubble horizon. Thus the horizon is crossed when $k c_s< aH$ (instead of $k < aH$), where $c_s$ is the speed of sound. This process lays the foundation for the creation of the large scale structure of the universe similar to that of standard inflation. \\

 While \cite{GarrigaandMukhanov, kinflation} use  vacuum fluctuations (or ground state) as the initial condition, \cite{Magueijo:2002pg} entertains the possibility that initial conditions were given by thermal fluctuations. In other words,  the universe starts near thermal equilibrium at a finite temperature $T$. 
%
In particular, \cite{Magueijo:2008pm} proposes a scenario, the so-called the Speedy Sound (or Tachyacoustic \cite{Kinneyetal}) cosmology, where the speed of sound rapidly decreases  from $c_s \gg 1$ in a very short interval of time. Depending on the acoustic and expansion histories, this could lead to scale-invariant perturbations, for either vacuum or thermal initial conditions (see also \cite{ArmendarizPicon:2003ht}). In these scenarios, allowing for superluminal propagation alleviates the need for cosmic inflation, at least in order to generate scalar scale-invariant perturbations on cosmological scales \cite{Geshnizjani:2011dk}. More exotic proposals for a thermal initial state in cosmology invoke the non-geometric non-locality inherent in string theory or quantum gravity \cite{Nayeri:2005ck,Magueijo:2006fu}.  

While superluminal propagation runs counter to the equivalence principle, which has been tested to extraordinary precison in a variety of circumstances (e.g. \cite{Liberati:2013xla}), it is proposed as a cure to high energy divergences of quantum field theory and quantum gravity. For example, in Horava-Lifshitz gravity \cite{Horava:2009uw}, a dispersion relation of $\omega \propto k^3$, which has an arbitrarily large speed of propagation at high energies, can lead to a power-counting renormalizable theory of gravity. Therefore, superluminal acoustic waves can be expected in such theories at sufficiently high temperatures.     


    
 
In this paper, we examine how the thermal tachyacoustic cosmology (TTC), first introduced in \cite{Magueijo:2008pm}, fares against current cosmological observations.    
Let us start by parametrizing the cosmological background that are phenomenologically consistent with our requirements.

We shall restrict our analyses to perturbations around flat FRW cosmology, given by the metric:
\beq
ds^2= a(\eta)^2 (d\eta^2 - d {\bf x} \cdot d{\bf x}) = dt^2- a^2 d {\bf x} \cdot d{\bf x},
\eeq
where $\eta$  and $t$ are the conformal and proper times, respectively, while $a$ is the cosmological scale factor. Furthermore, given the success of the hot big bang scenario for the early universe, we shall assume a radiation era expansion history, i.e. $a \propto \eta \propto t^{1/2}$, which implies $H = (2 t)^{-1}$. We shall further assume that the speed of sound decays as:
\beq
c_s = c_* \left(a\over a_*\right)^{-\beta} = c_* \left(\eta\over \eta_*\right)^{-\beta} ,
\eeq    
where $c_* \gg1 $ is the speed of sound at the moment of phase transition from the thermal equilibrium, to the radiation era, which happens at $a=a_*$, or $\eta = \eta_*$. Following \cite{Magueijo:2008pm}, we anticipate that $\beta \gg 1$ will be required to get nearly scale-invariant spectrum of scalar perturbations from thermal initial conditions.

The FRW metric with scalar linear perturbations in the longitudinal gauge is given by:
\beq
ds^2 = a^2\left[(1+2\Phi)d\eta^2 -(1-2\Psi)  d {\bf x} \cdot d{\bf x}\right]. 
\eeq
Observational constraints on scalar adiabatic perturbations are often described in terms of the gauge-invariant Bardeen variable \cite{Mukhanov:1990me}:
\beq
\zeta \equiv \Psi -\frac{H}{\dot{H}} (H\Phi +\dot{\Psi}). 
\eeq

Following \cite{GarrigaandMukhanov}, we shall adopt the following quadratic action for the Bardeen variable:
 \beq
 S= \frac{1}{2} M^2_{\rm p} \int dy d^3 {\rm x} ~q^2 \left[ \zeta'^2 - (\nabla \zeta)^2 \right],
 \eeq
 for acoustic waves of speed $c_s$, where
 \bea
 q &\equiv& \frac{a \sqrt{2\epsilon}}{\sqrt{c_s}} = \frac{2a}{\sqrt{c_*}} \left(a\over a_*\right)^{\beta/2} \propto \eta^{1+\beta/2},\label{q_def} \\
 y  &\equiv& \int \frac{c_s dt}{a} = \int c_s d\eta = -\frac{c_* \eta^\beta_*}{(\beta-1) \eta^{\beta -1}} \label{y_def} ,\\
 ' &\equiv & \frac{\partial}{\partial y},  ~\epsilon  \equiv  - \frac{\dot{H}}{H^2} =2\\
 M_{\rm p}  &\equiv&  (8\pi G_N)^{-1/2} = 2.435 \times 10^{18} ~{\rm GeV}.
\eea 
Here,  $y$ is the comoving sound horizon, and $M_{\rm p}$ is the reduced Planck mass. 

We can change to the Mukhanov-Sasaki variable:
\beq
v \equiv M_{\rm p} q \zeta,
\eeq
 which is canonically normalized:
\beq
 S= \frac{1}{2} \int dy d^3 {\rm x}  \left[v'^2 - (\nabla v)^2  + \frac{q''}{q} v^2 \right],
 \eeq
leading to the mode functions that obey the field equation in the Fourier space:               
\begin{equation}
v''_k + \left( k^2 - \frac{q''}{q} \right) v_k = 0.
\end{equation}
Combining Eqs. (\ref{q_def}) and (\ref{y_def}), we see that:
\beq
q \propto (-y)^{-\frac{\beta+2}{2\beta-2}} \Rightarrow \frac{q''}{q} = \frac{3\beta(\beta+2)}{4(\beta-1)^2y^2},
\eeq 
 and our mode equation becomes

\begin{equation}
v''_k + \left[ k^2 - \frac{3\beta(\beta+2)}{4(\beta-1)^2y^2} \right] v_k = 0. \label{mode_beta}
\end{equation}

So far, we have only considered the classical equations for linear perturbations. Following the standard canonical quantization procedure, we can decompose the free quantum fields in the Heisenberg picture as:

\begin{equation}
\hat{v}({\bf x},y) = \int{\frac{d^3 {\rm k}}{(2 \pi)^3} \left[ v_k(y) {\hat{a}}_{\bf k} e^{i {\bf k\cdot x} }
 + v_{k}^*(y) {\hat{a}}_{\bf k}^{\dagger} e^{-i {\bf k\cdot x}} \right]},  
\end{equation} 
where ${\hat{a}}_{\bf k}$ and $ {\hat{a}}_{\bf k}^{\dagger}$ are the creation and annihilation operators for particles (or phonons) of momentum ${\bf k}$ around a gaussian vacuum state $\left|0\right\rangle$, which, by definition, has zero particles. 

The adiabatic vacuum state $\left|0\right\rangle_{\rm ad.}$ is defined by the condition that mode functions $v_{k}(y)$ approach the positive frequency (flat space) limit, when $y\rightarrow -\infty$, which is also where adiabatic approximation in Eq. (\ref{mode_beta}) becomes exact:
\beq
v_{k}(y) \rightarrow \frac{\exp(-i ky)}{\sqrt{2k}}, {\rm ~when}~ y \rightarrow -\infty, \label{adiabatic}
\eeq
while its subsequent evolution follows from exactly solving the mode equation (\ref{mode_beta}). This ensures that the adiabatic vacuum coincides with the ground state of the Hamiltonian at infinite past. 

It turns out that Eq. (\ref{mode_beta}) with the initial condition (\ref{adiabatic}) has an exact solution in terms of the Hankel  function of the 2nd kind (or Bessel functions of 1st and 2nd kind):
\beq
v_{k}(y) = \frac{\sqrt{-\pi y}}{2} e^{-i\gamma_\nu}H^{(2)}_\nu (ky),\label{hankel}
\eeq
where 
\beq
\nu^2-\frac{1}{4} =  \frac{3\beta(\beta+2)}{4(\beta-1)^2}
\Rightarrow \nu = \frac{2\beta+1}{2\beta-2}, \gamma_\nu = \frac{\pi}{4}(2\nu+1). 
\eeq

The late-time power spectrum of $\zeta$ in a thermal state of temperature $T_*$ is given by:
\begin{equation}
\langle {\cal P}_{\zeta}(k)\rangle_{T_*} = \lim_{y \rightarrow 0^-} \frac{k^3}{2 {\pi}^{2}} \frac{{\left| v_k \right|}^2}{q^2M^2_{\rm p}} [2 \langle n_k\rangle_{T_*} + 1].\label{power}
\end{equation}
Here, the thermal particle occupation number is given by the Bose-Einstein distribution:
\beq
\langle n_k\rangle_{T_*} = \frac{1}{\exp\left(kc_*\over a_* T_*\right)-1}.\label{Bose-Einstein}
\eeq
We can also use the asymptotic form of Hankel function for small arguments:
\beq
\left| v_k \right|^2 = \frac{4^{\nu-1}\Gamma(\nu)^2}{\pi y^{2\nu-1} k^{2\nu}} + {\cal O} (y^{2-2\nu}). \label{asymptote}
\eeq

Combining Eqs. (\ref{power})-(\ref{asymptote}) yields:
\begin{widetext}
\bea
 \langle {\cal P}_{\zeta}(k)\rangle_{T_*} = \frac{[2(\beta-1)]^{1+\frac{3}{\beta-1}}(\eta_*/a_*)^2(c_*\eta_*^\beta)^{-\frac{3}{\beta-1}}}{16\pi^3 M^2_{\rm p}} \Gamma\left[1+\frac{3}{2(\beta-1)}\right]^2\left[ \frac{2}{\exp\left(kc_*\over a_* T_*\right)-1} +1\right] k^{1-\frac{3}{\beta-1}} \nonumber\\
 = \frac{[2(\beta-1)]^{1+\frac{3}{\beta-1}}}{16\pi^3}\Gamma\left[1+\frac{3}{2(\beta-1)}\right]^2 \left(H_*\over M_{\rm p}\right)^2 \left[ \frac{2}{\exp\left(kc_*\over a_* T_*\right)-1} +1\right] \left(k\over a_* H_*\right) \left(k c_*\over a_* H_*\right)^{-\frac{3}{\beta-1}}.\label{power_wide}
\eea
\end{widetext}

Notice that, in the Rayleigh-Jeans limit $kc_* \ll a_*T_*$, we have  $\langle n_k\rangle_{T_*} \propto k^{-1}$, and thus the scalar spectral index is given by:
\bea
n_s-1&=& -\frac{3}{\beta-1} = -0.0397 \pm
0.0073 \nonumber\\ &\Rightarrow &\boxed{ \beta = 77 ^{+17}_{-12}}\label{beta_box}
\eea
where we used the observational constraints on the scalar power spectrum from Eq. (\ref{p_zeta}). Indeed, as anticipated in \cite{Magueijo:2008pm}, a sufficiently rapid drop in speed of sound can yield a near scale-invariant power spectrum of scalar perturbations, starting from thermal initial conditions.  

Now, for $\beta \gg 1$, the amplitude of $P_\zeta$ is roughly given by:
\bea
&& \langle {\cal P}_{\zeta}(k)\rangle_{T_*}  \nonumber \\ &&\simeq \left[\beta +3\ln(2\beta) -3\gamma-1 \over 4\pi^3 \right] \left(\frac{H_*T_*}{M^2_{\rm p}}\right) \left(a_*\over a_{\rm exit}(k) \right)^3 \nonumber\\ &&= \left[\beta +3\ln(2\beta) -3\gamma-1 \over 4\pi^3\right] \left(\frac{H_{\rm exit}(k)T_{\rm exit}(k)}{ M^2_{\rm p}}\right),
\eea
where $\gamma \simeq 0.577216$ is the Euler's constant, while the subscript ``exit'' refers to when the mode exits the sound horizon, i.e. 
\beq
k c_s(a_{\rm exit}) =a_{\rm exit}H_{\rm exit}.
\eeq
Now, using Friedmann equation in the radiation era:
\beq
H^2=  \frac{\pi^2}{90} \frac{g_* T^4}{M^2_{\rm p}},\label{fried}
\eeq
where $g_*$ is the effective number of relativistic species, we can express amplitude in terms of temperature at sound horizon exit:
\beq
 \langle {\cal P}_{\zeta}(k)\rangle_{T_*}  \simeq  2.4 \left(\beta \over 77\right) \left( g_* \over 100\right)^{1/2} \left[ T_{\rm exit} (k) \over M_{\rm p}\right]^{3},
\eeq
which, comparing to the Planck constraint on the primordial power spectrum (\ref{p_zeta}), yields:
\begin{empheq}[box=\fbox]{align}
&T_{\rm exit} (k=0.05~ {\rm Mpc}^{-1}) \simeq \nonumber\\ &  \left( g_* \over 100\right)^{-1/6} (2.368 \pm 0.021 \pm 0.154) \times 10^{15}   {\rm GeV}, \label{texit}
\end{empheq}
where the first error comes from the measurement uncertainty on power spectrum amplitude (\ref{p_zeta}), while the second error comes from uncertainty in $\beta$ (Eq. \ref{beta_box}), which in turn depends on the uncertainty in the measurement of scalar spectral index in (\ref{p_zeta}). 

\begin{figure}
\includegraphics[scale = 0.8]{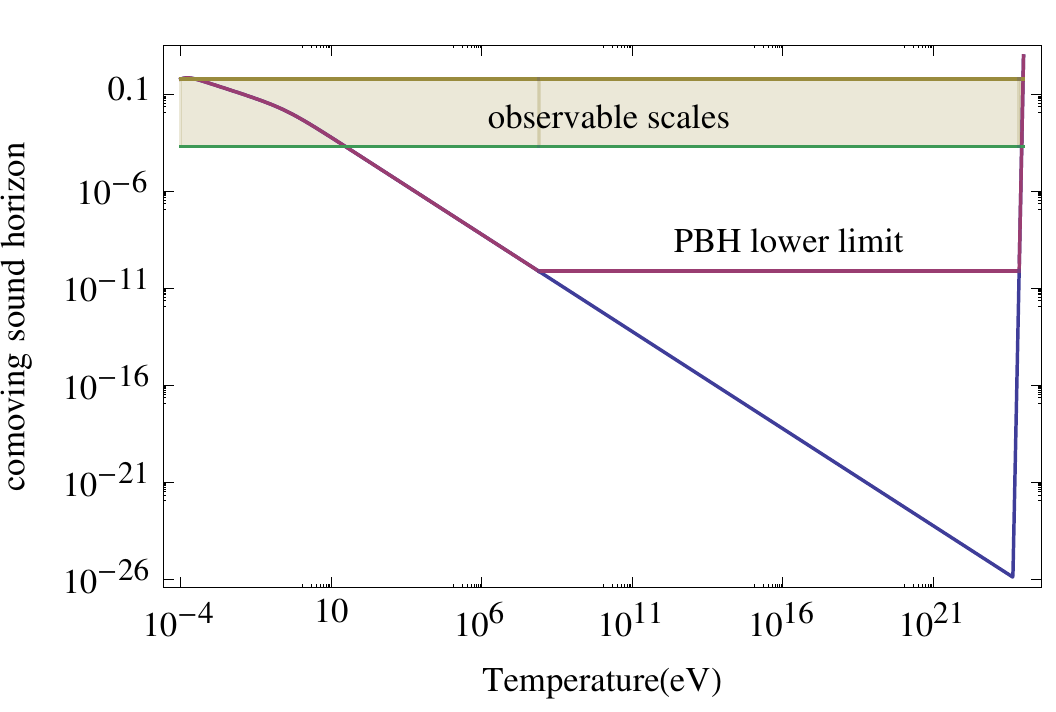}
\caption{Evolution of the comoving sound horizon $c_s/(aH)$, or {\it acoustic history}, in Thermal Tachyacoustic Cosmology (TTC). The shaded region shows the observable comoving scales, which all exit the sound horizon at $T_{\rm exit} \sim 10^{15} ~{\rm GeV} = 10^{24} ~{\rm eV} $. The horizontal line shows the rough lower limit on the comoving sound horizon, to avoid a catastrophic production of primordial black holes (PBH's). \label{sound}}
\end{figure}

Fig. (\ref{sound}) shows the approximate evolution of sound horizon, or {\it acoustic history},  in TTC, roughly assuming that $T\propto a^{-1}$. Since the horizon problem is solved by the sudden drop of the sound speed,  the speed of sound at phase transition must be $c_* \gtrsim 10^{25} \sim e^{58} $, which is similar to the $\sim$ 60 e-foldings required in cosmic inflation. However, the requirement of near-scale-invariance of the scalar power spectrum, which led to $\beta \sim 80$, implies that this sudden drop from $c_s \sim 10^{25}$ to $c_s \sim 1$ should happen in less than a Hubble time! Moreover, similar to inflation, the characteristic temperature for the phase transition is the GUT scale, $T\sim 10^{15}$ GeV.   

Let us now address some of the phenomenological issues that may arise in TTC:

We first should note that, in order to get a scale-invariant power spectrum (\ref{beta_box}), we had to assume to be in the Rayleigh-Jeans limit of the Bose-Einstein distribution. As we approach the thermal wavenumber $kc_* \sim a_* T_*$, the vacuum fluctuation overtake the thermal fluctuations in Eq. (\ref{power_wide}), which would in turn lead to a blue spectrum $n_s \simeq 2$ for $\beta \gg 1$. This could lead to curvature fluctuations of $\zeta \sim 1$, on sufficiently small scales, leading to a catastrophic production of primordial black holes (PBH's) upon horizon re-entry, which could prematurely end the radiation era (i.e. overclose the universe). In order to avoid overproducing PBH's, we need to ensure that  the modes with ${\cal P}_\zeta \gtrsim 0.05$ never exit the sound horizon \cite{Young:2014ana}.  In the $kc_* > a_* T_*$ (or Wien) regime, this implies:
\beq
{\cal P}_\zeta \sim  \frac{1}{2}{\cal P}_{\zeta,{\rm  IR}} \left(a_{\rm exit} H_{\rm exit} c_* \over c_{\rm exit} a_* T_*\right) \lesssim 0.05,
\eeq 
where $ {\cal P}_{\zeta, {\rm IR}}  \sim 2\times 10^{-9}$ is the observed spectrum of curvature perturbations. This implies:
\bea
\frac{c_{\rm exit}}{a_{\rm exit} H_{\rm exit}} \gtrsim 2\times 10^{-8} \left(H_*\over T_*\right) \left(\frac{c_*}{a_* H_*}\right) \nonumber\\ \gtrsim 6 \times 10^{-11}  \left( g_* \over 100\right)^{1/3}  \left(\frac{c_*}{a_* H_*}\right),
\eea
where we used $T_{\rm exit} < T_*$ and the Friedmann equation (\ref{fried}) in the last step. This lower bound on the comoving sound horizon is also shown in Fig. (\ref{sound}), suggesting that the sound speed should drop rapidly by $\sim 10$ orders of magnitude, within a fraction of Hubble time when $T \sim T_{\rm GUT} \sim 10^{15}$ GeV, but then gradually decrease by a further $\sim 15$ orders of magnitude, from GUT to QCD phase transition ($T_{\rm QCD} \sim 100$ MeV), approaching the relativistic value just in time for the big bang nucleosynthesis. 

We can also make an interesting prediction for the amplitude of primordial gravitational waves (or tensor modes), if we relate the phase transition in TTC to Horava-Lifshitz gravity \cite{Horava:2009uw}. Using Friedmann equation, we can estimate the Hubble constant at the time/temperature of TTC phase transition (Eq. \ref{texit}):
\beq
H_* \gtrsim H_{\rm exit}(k=0.05~ {\rm Mpc}^{-1}) \sim 8\times 10^{12} \left( g_* \over 100\right)^{1/6}  {\rm GeV}. \label{hstar}
\eeq
If the phase transition is triggered by Lifshitz symmetric operators in the gravity theory, then one expects the Lifshitz scale to be comparable to the expansion rate at phase transition, i.e. $\Lambda_{\rm HL} \sim H_*$. Meanwhile, since $\omega \sim k^3/\Lambda_{\rm HL}^2$ for gravitons in Horava-Lifshitz theory, the vacuum spectrum of gravitational waves is scale-invariant:
\beq
{\cal P}_T \simeq \frac{2\Lambda_{HL}^2}{\pi^2 M^2_{\rm p}} \sim   \frac{2H_*^2}{\pi^2 M^2_{\rm p}}, \label{ptensor}
\eeq
which is, incidentally, similar to the inflationary prediction. 
Now, combining Eqs. (\ref{p_zeta}), (\ref{hstar}), and (\ref{ptensor}) yields:
\beq
\boxed{r \equiv \frac{{\cal P}_T}{{\cal P}_\zeta} \gtrsim 1 \times 10^{-3}  \left( g_* \over 100\right)^{1/3}, }
\eeq
i.e. the tensor-to-scalar ratio must be bigger than $\sim 10^{-3}$ in TTC coupled to Horava-Lifshitz gravity, which is potentially detectable by CMBpol type satellite polarization experiments \cite{Baumann:2008aq}. Optimistically,  it is tantalizing to entertain that this effect could be responsible for (the primordial part of) the signal detected by BICEP2 experiment \cite{Ade:2014xna} (assuming both our predictions and BICEP2 observations stand the test of time). 

Let us summarize our results: We have studied an intriguing alternative to inflation, Thermal Tachyacoustic Cosmology (TTC), which could explain  observed  cosmological initial conditions by employing an initial phase of thermal equilibrium, with extremely superluminal acoustics, followed by a rapid phase transition into the radiation era. This scenario fits well into Lorentz-violating  attempts at UV-completion of gravity, such as in Horava-Lifshitz proposal. We also derived constraints on the theory necessary to fit cosmological observations, which amount to a very rapid phase transition around the GUT scale. However, in order to avoid catastrophic production of  PBH's, a secondary slower phase is necessary. Finally, we argued that coupling TTC to Horava-Lifshitz gravity puts a lower bound on the tensor-to-scalar ratio $\gtrsim 10^{-3}$, that could be detectable by future experiments, if not already detected by the BICEP2 collaboration. As a next step, it will be imperative to come up with a concrete action, {\it beyond the quadratic level}, that could reproduce the effective behaviours outlined here, and can be used to make further testable predictions for the TTC. 

AA wishes to thank Eileen Maeve Manion Fischer, Kai Cheong Chan and Mukto Akash for useful discussions.  NA is supported by the Natural Science and Engineering Research Council of Canada, the University of Waterloo, and Perimeter Institute for Theoretical Physics. Research
at Perimeter Institute is supported by the Government of Canada through Industry
Canada and by the Province of Ontario through the Ministry of Research \& Innovation.

\bibliography{Bibliography.bib}

\end{document}